\documentclass[aps,showpacs,prl,superscriptaddress,twocolumn,secnumarabic,balancelastpage,nofootinbib,hyperref=pdftex,titlepage] {revtex4}
\usepackage{graphicx}  
\usepackage{amsmath} 
\usepackage{amssymb}   
\usepackage{siunitx}
\usepackage[colorlinks=true]{hyperref}

\newcommand{\ket}[1]{\left| #1 \right>} 
\newcommand{\bra}[1]{\left< #1 \right|} 
\renewcommand{\vec}[1]{\mathbf{#1}} 

\begin{document}
\title{Modified hyper-Ramsey methods for the elimination of probe shifts in optical clocks}
\author{R. Hobson}
	\affiliation{National Physical Laboratory, Hampton Road, Teddington, TW11 0LW, UK}
	\affiliation{Clarendon Laboratory, Parks Road, Oxford OX1 3PU, UK}
\author{W. Bowden}
	\affiliation{National Physical Laboratory, Hampton Road, Teddington, TW11 0LW, UK}
	\affiliation{Clarendon Laboratory, Parks Road, Oxford OX1 3PU, UK}
\author{S. A. King}
	\affiliation{National Physical Laboratory, Hampton Road, Teddington, TW11 0LW, UK}
\author{P. E. G. Baird}
	\affiliation{Clarendon Laboratory, Parks Road, Oxford OX1 3PU, UK}
\author{I. R. Hill}
	\affiliation{National Physical Laboratory, Hampton Road, Teddington, TW11 0LW, UK}
\author{P. Gill}
	\affiliation{National Physical Laboratory, Hampton Road, Teddington, TW11 0LW, UK}

\begin{abstract}
We develop a method of modified hyper-Ramsey spectroscopy in optical clocks, achieving complete immunity to the frequency shifts induced by the probing fields themselves. Using particular pulse sequences with tailored phases, frequencies, and durations, we can derive an error signal centered exactly at the unperturbed atomic resonance with a steep discriminant which is robust against variations in the probe shift. We experimentally investigate the scheme using the magnetically-induced $^1$S$_0- ^3$P$_0$ transition in $^{88}$Sr, demonstrating automatic suppression of a sizeable \num{2e-13} probe Stark shift to below \num{1e-16} even with very large errors in shift compensation.
\end{abstract}


\maketitle

High-$Q$ interrogation of narrow, forbidden optical transitions has formed the basis of a new generation of atomic clocks with exceptional accuracy and stability at the $10^{-18}$ level \cite{Hinkley2013, Nicholson2015}. As well as potentially supporting a redefinition of the SI second \cite{Gill2011}, such clocks underpin empirical investigations into areas of fundamental physics including relativity \cite{Chou2010}, the search for dark matter \cite{Derevianko2014,VanTilburg2015}, and potential time-variation of fundamental constants \cite{Godun2014,Huntemann2014}. However, several promising atomic species suffer from a clock transition which is too forbidden, requiring a high laser intensity $\--$ and therefore a large light shift $\--$ in order to drive the atoms into the excited state. Important examples include clocks based on high-order multipole \cite{Huntemann2012,King2012}, two-photon \cite{Zanon-Willette2006,Parthey2011}, or magnetically-induced \cite{Taichenachev2006a, Barber2006, Akatsuka2010} transitions. Before these species can be used for precision frequency measurements, probe-induced shifts must be dealt with.

A conceptually straightforward approach to the light shift is exemplified by recent realizations of the electric-octupole $^2$S$_{1/2}$ $\rightarrow$ $^2$F$_{7/2}$ $^{171}$Yb$^+$ clock, where the unperturbed atomic resonance is extrapolated from interleaved Rabi-spectroscopy sequences of high and low probe intensity \cite{Godun2014,Huntemann2012}. Although careful extrapolation can be quite effective, achieving frequency uncertainty nearly 4 orders of magnitude smaller than the shift itself \cite{Godun2014}, the stability of the clock is deteriorated by the extrapolation process and the ultimate accuracy is limited by the precision with which the probe intensity ratio can be calibrated.

In order to reduce the burden of probe intensity control, tailored spectroscopy pulses have been proposed which provide a central feature whose frequency is unchanged by the light shift \cite{Taichenachev2010,Yudin2010,Zanon-Willette2014,Zanon-Willette2015}. The great potential of this approach has recently been illustrated using Yb$^+$, where ``hyper-Ramsey'' spectroscopy was utilized to suppress the shift by more than four orders of magnitude below the $10^{-17}$ level \cite{Huntemann2012a}. In this Letter, we propose a modified form of hyper-Ramsey spectroscopy which provides complete immunity to variations in the probe shift, considerably relaxing the experimental constraints on intensity control and potentially facilitating light shift uncertainties well below $10^{-18}$ in Yb$^+$. As indicated by the experimental results of this Letter, sub-$10^{-18}$ shift uncertainty is also made accessible for $^{88}$Sr clocks based on the magnetically-induced $^1$S$_0$ $\rightarrow$ $^3$P$_0$ transition. Our method smooths the path to high accuracy for clocks using less stable local oscillators where shorter, more intense pulses are a necessity, and also improves the prospects for state-of-the-art frequency measurements with direct frequency-comb spectroscopy \cite{Fortier2006} or using highly-charged ions with ultra-forbidden clock transitions \cite{Safronova2014}.

\begin{figure}[b]
\begin{center}
\includegraphics[width=8.6cm]{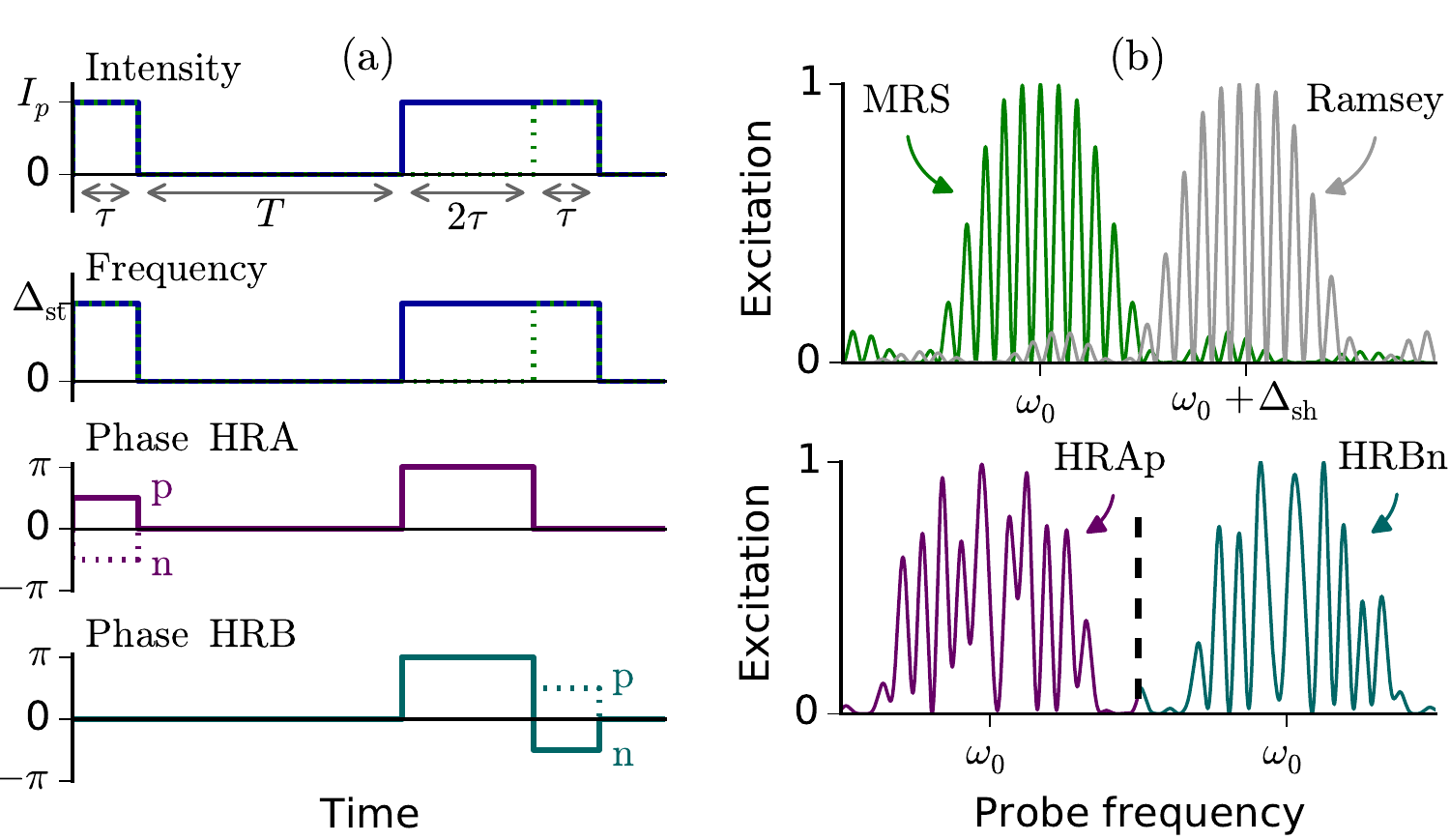}
\end{center}
\caption{(color online). Pulse patterns (a) and excitation spectra (b) of the various spectroscopy methods considered in this paper. In the top half of (a), the blue traces indicate the common intensity and frequency patterns of hyper-Ramsey types A and B, while the dotted green traces show modified Ramsey spectroscopy (MRS). The spectra in (b) are calculated with $T = 4\tau$, $\Omega_0\tau=\pi/2$, $\Delta_\mathrm{sh}/2\pi = 1.56/\tau$, and $\Delta_\mathrm{st} = \Delta_\mathrm{sh}$.}
\label{fig:Hyper-Ramsey_schematics}
\end{figure}

Taking the Ramsey method of separated oscillatory fields \cite{Ramsey1950} as the starting point for our discussion of these probe-immune spectroscopy techniques, we observe two problems which emerge when a large probe shift is present: the envelope of the Ramsey fringes moves to follow the shift, and the distribution of fringes themselves becomes asymmetric around the peak (see figure \ref{fig:Hyper-Ramsey_schematics}). A solution is pointed out in \cite{Taichenachev2010}, which we call ``modified Ramsey'' spectroscopy (MRS): if we apply a frequency step $\Delta_\mathrm{st}$ to the probe laser to compensate exactly for the shift $\Delta_\mathrm{sh}$, then both the envelope and the Ramsey fringes are restored exactly back to the position of the perfect, unperturbed Ramsey spectrum. However, applying the correct frequency step requires us to know exactly what the light shift is, which is experimentally unrealistic. When the compensation step is slightly incorrect $\--$ i.e. when a residual uncompensated shift $\Delta = \Delta_\mathrm{sh} - \Delta_\mathrm{st}$ remains $\--$ the frequency of the central MRS fringe moves, feeding directly into a frequency shift of the locked clock.

To reduce sensitivity to $\Delta$, it was proposed in \cite{Yudin2010} to insert an extra pulse into the Ramsey dark time. In detail, this proposal corresponds to the ``hyper-Ramsey type A'' spectroscopy (HRA) depicted in figure \ref{fig:Hyper-Ramsey_schematics}: using a constant laser intensity during the pulses, we expose the atoms to a pulse of length $\tau$ and phase $\pm \pi/2$, then wait for dark time $T$, and finally apply a pair of pulses of length $2\tau$, $\tau$ and phases $\pi$, $0$ respectively. The intensity should ideally be chosen to give a first pulse area close to $\pi/2$ to maximize the frequency discriminant, but errors in pulse area have no significant effect on the probe shift suppression \cite{Yudin2010,Zanon-Willette2015}. The atomic clock is stabilized to the hyper-Ramsey feature as follows: Interleaving one sequence of HRA$_\mathrm{p}$ (phase $+\pi/2$) followed by one sequence of HRA$_\mathrm{n}$ (phase $-\pi/2$) we use the difference in excitation fractions as an error signal to feedback to the clock frequency. In the long term, the clock frequency will settle at the point where the two excitation probabilities $P$ are the same, i.e. where $P_\mathrm{HRA_p} = P_\mathrm{HRA_n}$.

\begin{figure}
\begin{center}
\includegraphics[width=8cm]{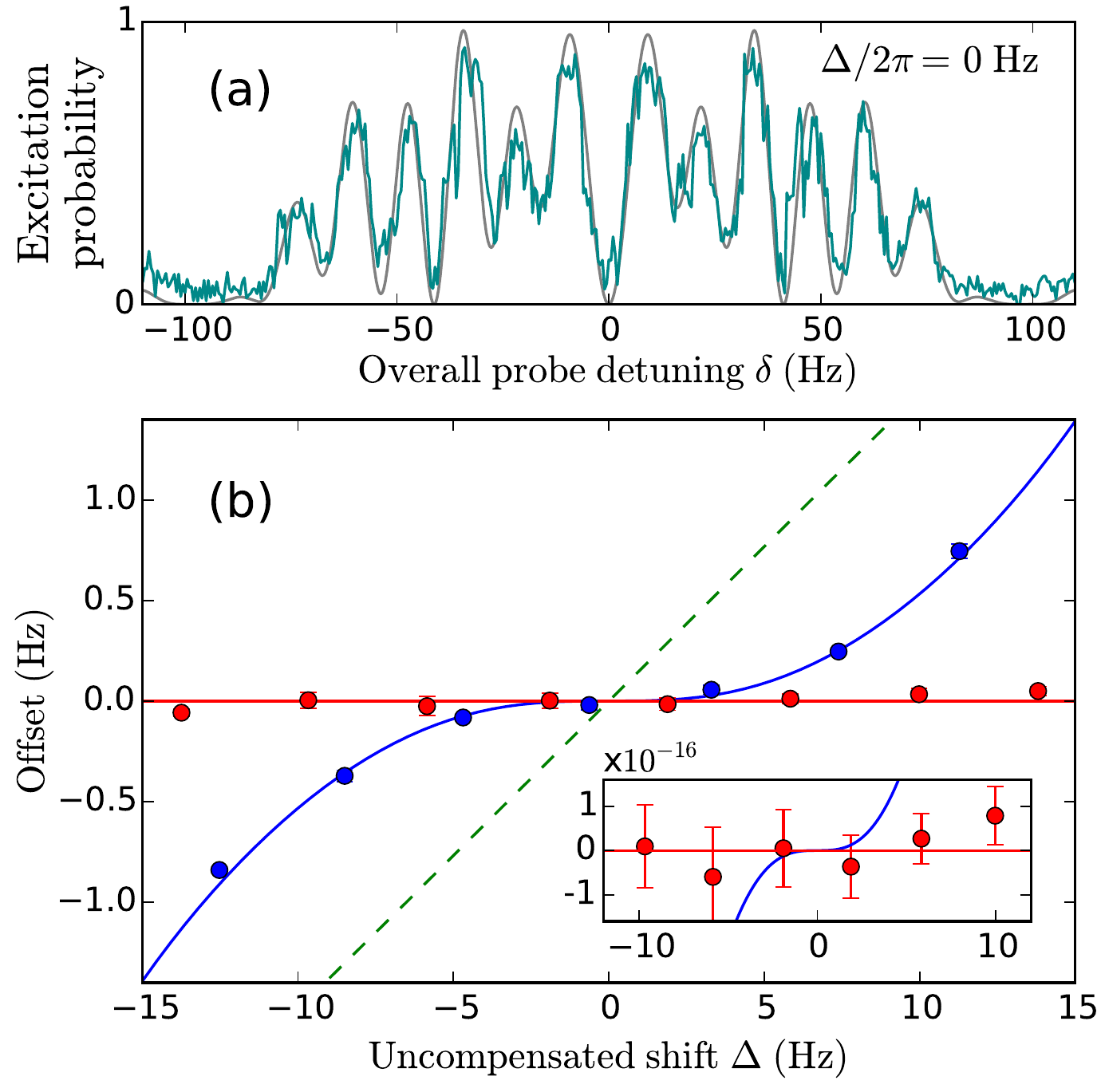}
\end{center}
\caption{(color online). (a) Scan over the hyper-Ramsey feature in $^{88}$Sr for $\tau = \SI{10}{\milli\second}$, $T = \SI{50}{\milli\second}$, and $\Omega_0\tau \approx \pi/2$, and with $\phi = 0$ in the first and last pulse. The theoretical model is overlaid in gray with no fitting parameters used. (b) Modeled and measured residual Stark shifts for the different spectroscopy methods: The $\mathrm{HRA_p}$ \& $\mathrm{HRA_n}$ `standard' hyper-Ramsey lock (blue) shows good suppression compared with modified Ramsey (dashed green), but the $\mathrm{HRA_n}$ \& $\mathrm{HRB_p}$  modified hyper-Ramsey (red) is better. \textit{Inset:} Enlarged view showing the residual lock offset in fractional frequency units.}
\label{fig:Hyper-Ramsey_shift_curve}
\end{figure}

Using the HRA spectroscopy pattern, not only is the locked frequency equal to the unperturbed $\omega_0$ when we apply exactly the correct compensation step, but its linear dependence on small variations in $\Delta$ is also eliminated \cite{Yudin2010}. However, the HRA clock remains vulnerable to a residual cubic dependence on $\Delta$ (see figure \ref{fig:Hyper-Ramsey_shift_curve}), thus still requiring careful control of the compensation step $\Delta_\mathrm{st}$ to within a small region around $\Delta_\mathrm{sh}$ to avoid significant shifts. In this paper we introduce the hyper-Ramsey `type B' (HRB) spectroscopy depicted in figure \ref{fig:Hyper-Ramsey_schematics}, which is identical to HRA except that the $\pm \pi/2$ phase step is implemented in the last pulse instead of the first. Using a mix of both HRA and HRB spectroscopy we can ultimately eliminate all dependence of the locked clock on $\Delta$, providing a clock lockpoint at exactly $\omega_0$ regardless of errors in the compensation step.

In the rotating frame of the laser field $\vec{E}(t) = \vec{E}_0 \cos((\omega_L+\Delta_\mathrm{st}) t + \phi)$, we model the evolution of the atomic state during the probe pulses in the basis $\ket{g} = \big(\begin{smallmatrix} 0 \\ 1 \end{smallmatrix}\big)$ and $\ket{e} = \big(\begin{smallmatrix} 1 \\ 0 \end{smallmatrix}\big)$ using the propagator:

\begin{align}
& \hat{W}(t_p,\Omega_0,\Delta_p,\phi) \nonumber \\
&= \begin{pmatrix} \cos(\frac{\Omega t_p}{2}) + i\frac{\Delta_p}{\Omega}\sin(\frac{\Omega t_p}{2}) & - i e^{-i\phi}\frac{\Omega_0}{\Omega}\sin(\frac{\Omega t_p}{2}) \vphantom{\bigg)} \\ -ie^{i\phi} \frac{\Omega_0}{\Omega}\sin(\frac{\Omega t_p}{2}) & \cos(\frac{\Omega t_p}{2}) - i\frac{\Delta_p}{\Omega}\sin(\frac{\Omega t_p}{2}) \end{pmatrix} \label{eq:general_atom_evolution}
\end{align}
where we define a probe time $t_p$, a Rabi frequency $\Omega_0 = \vec{d}\cdot\vec{E}/\hbar$, a generalized Rabi frequency $\Omega = \sqrt{\Omega_0^2 + \Delta_p^2}$, and an effective laser detuning $\Delta_p = \omega_L - \omega_0 - \Delta$.

Meanwhile, the propagator for the hyper-Ramsey dark time $T$ is given by:

\begin{equation}
\hat{V}(T,\delta) = \begin{pmatrix} e^{i\frac{\delta T}{2}} & 0 \vphantom{\bigg)}\\ 0 & e^{-i\frac{\delta T}{2}} \end{pmatrix}
\end{equation}
where $\delta = \omega_L - \omega_0$.

For the example of hyper-Ramsey spectroscopy of type A with a positive $\pi/2$ phase step, we can then write the total propagator for the sequence as:

\begin{align}
\hat{U}_\mathrm{HRA_p} = \hat{W}(\tau,\Delta_p,0)\hat{W}(2\tau,\Delta_p,\pi)\hat{V}(T,\delta)\hat{W}(\tau,\Delta_p,\pi/2)
\end{align}
while propagators for the other three types of hyper-Ramsey look identical except for different phases in the first and last pulse.

\begin{figure}
\begin{center}
\includegraphics[width=8cm]{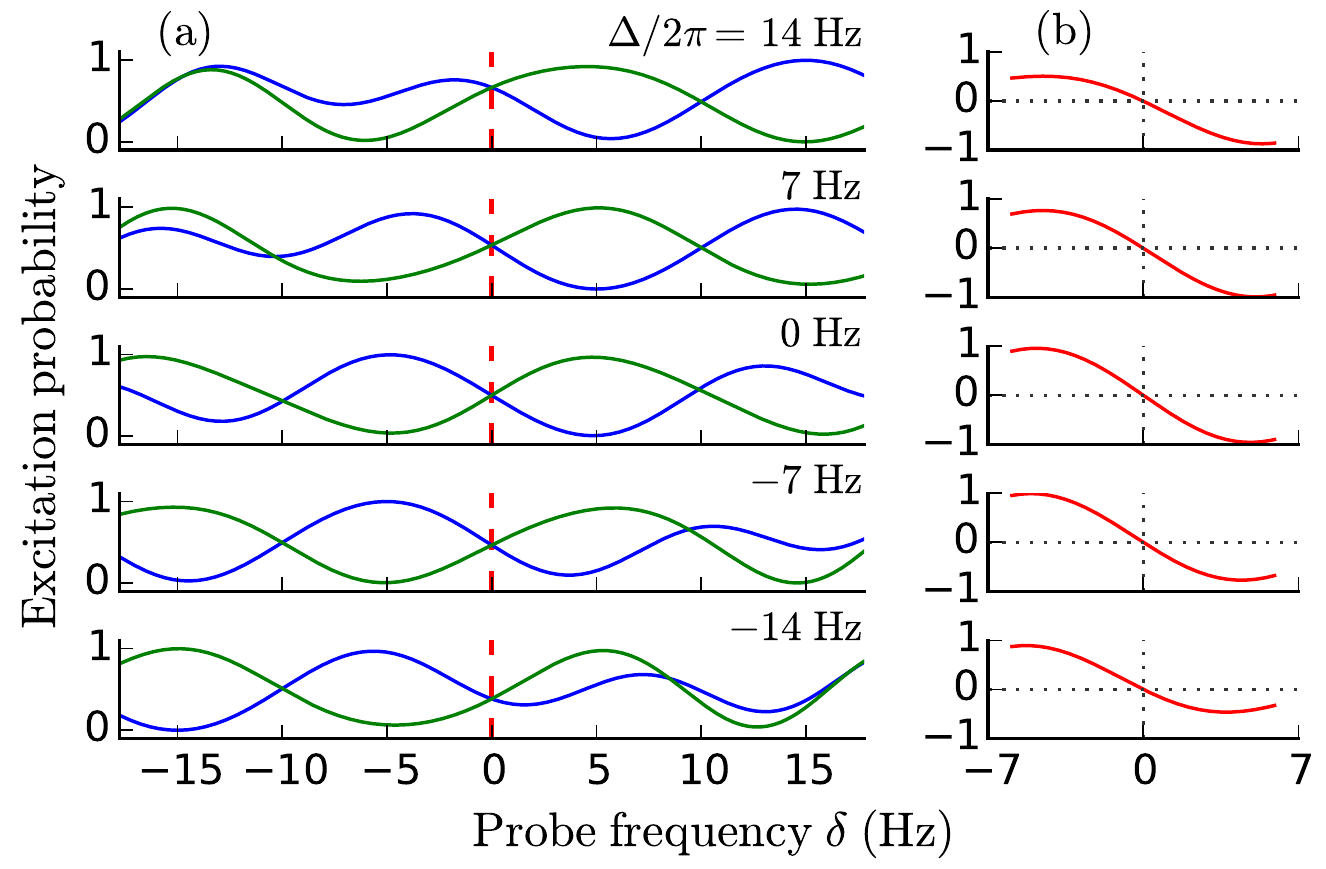}
\end{center}
\caption{(color online). (a) Excitation spectra and (b) error signals $P_\mathrm{HRA_p} - P_\mathrm{HRB_n}$ around the central hyper-Ramsey fringe for different uncompensated shifts. In (a), the blue and green curves correspond to $\mathrm{HRA_p}$ and $\mathrm{HRB_n}$ sequences respectively and the red dashed lines highlight that the excitation probabilities always cross at $\delta=0$. We use $\Omega_0\tau=\pi/2$, $\tau=\SI{10}{\milli\second}$, and $T=\SI{50}{\milli\second}$.}
\label{fig:Hyper_Ramsey_central_fringe_vs_uncomp_shift}
\end{figure}

We observe a set of important symmetries for the evolution during the pulses (equation \ref{eq:general_atom_evolution}):

\begin{align}
\hat{W}(t_p,\Omega_0,\Delta_p,\phi)^\dagger &= \hat{W}(t_p,\Omega_0,-\Delta_p,\phi+\pi) \label{eq:hermitian_conj_sym}\\
\hat{W}\left(t_p,\Omega_0,\Delta_p,(0,\pi)\right)^T &= \hat{W}\left(t_p,\Omega_0,\Delta_p,(0,\pi)\right) \label{eq:transpose_sym_0_pi} \\
\hat{W}\left(t_p,\Omega_0,\Delta_p,\pm\pi/2\right)^T &= \hat{W}\left(t_p,\Omega_0,\Delta_p,\mp\pi/2\right) \label{eq:transpose_sym_pi_by_2}
\end{align}

If we consider hyper-Ramsey sequences where the laser is on resonance such that $\delta = 0$, $\Delta_p = \Delta$ and $\hat{V}(T,\delta)=\hat{\mathbb{I}}$, we can use the symmetry eqns \ref{eq:hermitian_conj_sym} - \ref{eq:transpose_sym_pi_by_2} to derive general relationships between the propagators for different types of hyper-Ramsey interrogation. For instance:

\begin{align}
\hat{U}_{\delta=0,\mathrm{HRA_p}}^T &= \left(\hat{W}(\tau,\Delta,0)\hat{W}(2\tau,\Delta,\pi)\hat{W}(\tau,\Delta,\pi/2)\right)^T \nonumber\\
&= \hat{W}(\tau,\Delta,\pi/2)^T \hat{W}(2\tau,\Delta,\pi)^T \hat{W}(\tau,\Delta,0)^T \nonumber \\
&= \hat{W}\left(\tau,\Delta,-\pi/2\right) \hat{W}(2\tau,\Delta,\pi) \hat{W}(\tau,\Delta,0) \nonumber \\
&= \hat{U}_{\delta=0,\mathrm{HRB_n}}
\end{align}
and similarly $\hat{U}^T_{\delta=0,\mathrm{HRA_n}} = \hat{U}_{\delta=0,\mathrm{HRB_p}}$.

Using the fact that all the total propagators must be unitary, these results imply equalities between the on-resonance excitation probabilities:

\begin{align}
P_\mathrm{HRA_p} &= |\bra{e}\hat{U}^T_{\delta=0,\mathrm{HRA_p}} \ket{g}|^2 \nonumber\\
&= |\bra{e}\hat{U}_{\delta=0,\mathrm{HRB_n}} \ket{g}|^2 = P_\mathrm{HRB_n} \label{eq:equal_exc}
\end{align}
and similarly $P_\mathrm{HRA_n} = P_\mathrm{HRB_p}$.

The effect of the equality in equation \ref{eq:equal_exc} is illustrated in figures \ref{fig:Hyper_Ramsey_central_fringe_vs_uncomp_shift} and \ref{fig:Hyper-Ramsey_shift_curve}. Over a range of different uncompensated shifts $\Delta$, we see that the hyper-Ramsey fringes always cross at the unperturbed atomic resonance $\omega_L=\omega_0$. This represents the critical result of our scheme: Using an error signal proportional to either $P_\mathrm{HRA_p} - P_\mathrm{HRB_n}$ or $P_\mathrm{HRB_p} - P_\mathrm{HRA_n}$ we realize a locked clock laser frequency at $\omega_0$ which is immune to variations in $\Delta$. Note also that this result is independent of pulse area $\Omega_0\tau$, meaning that exact realization of a $\pi/2$ pulse area is not needed.

However, although the lineshape crossing is robust against $\Delta$, figure \ref{fig:Hyper_Ramsey_central_fringe_vs_uncomp_shift} also reveals that the slope at the crossing is reduced if the uncompensated shift is too large. If this reduced slope were allowed to persist, the stability of the locked clock would be compromised and its accuracy could be contaminated by asymmetries in the frequency discriminant around the zero-crossing. Therefore, even in the modified hyper-Ramsey scheme we still need to make sure that the shift compensation is roughly correct. It would be possible to interrupt the clock operation with a separate Rabi servo to evaluate the probe shift over time (as proposed in \cite{Huntemann2012a}), but another useful symmetry between the different types of hyper-Ramsey spectroscopy provides us with a method of servoing $\Delta$ towards 0 without degrading clock stability. In particular, we have a $\delta=0$ evolution:

\begin{align}
\hat{U}_{\delta=0,\mathrm{HRA_p}}^\dagger &= \left(\hat{W}(\tau,\Delta,0)\hat{W}(2\tau,\Delta,\pi)\hat{W}(\tau,\Delta,\pi/2)\right)^\dagger \nonumber\\
&= \hat{W}(\tau,\Delta,\pi/2)^\dagger \hat{W}(2\tau,\Delta,\pi)^\dagger \hat{W}(\tau,\Delta,0)^\dagger \nonumber \\
&= \hat{W}\left(\tau,-\Delta,-\pi/2\right) \hat{W}(2\tau,-\Delta,0) \hat{W}(\tau,-\Delta,\pi) \nonumber \\
&= \hat{W}\left(\tau,-\Delta,\pi/2\right) \hat{W}(2\tau,-\Delta,\pi) \hat{W}(\tau,-\Delta,0) \nonumber \\
&= \hat{U}_{\delta=0,\mathrm{HRB_p}}(-\Delta)
\end{align}
where in the penultimate step we have added a global phase $\pi$ to all the pulses which will not affect the atom dynamics. This implies further equalities between on-resonance excitation probabilities:

\begin{align}
P_\mathrm{HRA_p}(\Delta) = P_\mathrm{HRB_n}(\Delta) = P_\mathrm{HRA_n}(-\Delta) = P_\mathrm{HRB_p}(-\Delta)
\end{align}

\begin{figure}
\begin{center}
\includegraphics[width=8cm]{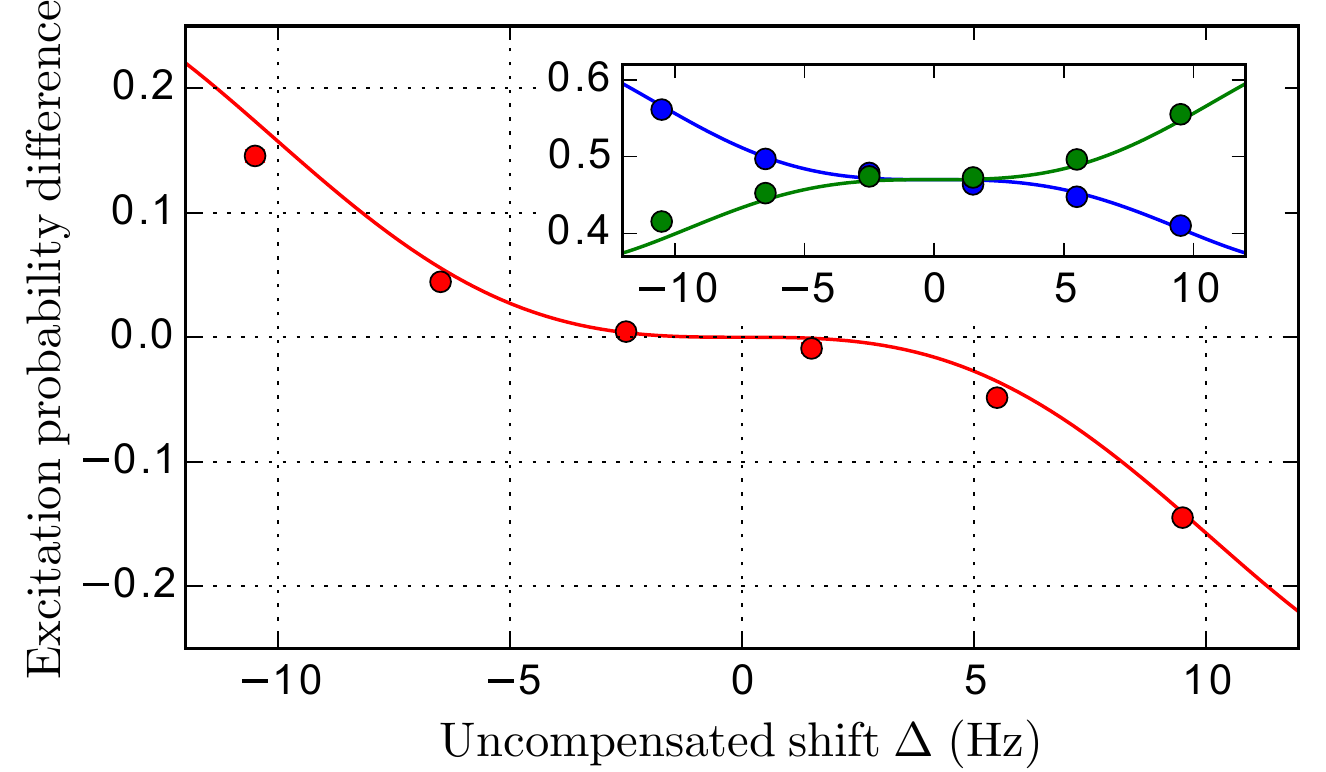}
\end{center}
\caption{(color online). \textit{Inset:} The dependencies of equilibrium $\delta=0$ excitation fraction on uncompensated shift for the both modified hyper-Ramsey locks: $\mathrm{HRA_p}$ \& $\mathrm{HRB_n}$ (green) and $\mathrm{HRA_n}$ \& $\mathrm{HRB_p}$ (blue). \textit{Main:} Excitation difference to be used for steering the compensation step towards $\Delta = 0$. The model is scaled by 0.94 to match the experimental fringe contrast.}
\label{fig:Excitation-based_step_servo_signal}
\end{figure}

The results of this equality are illustrated in the inset of figure \ref{fig:Excitation-based_step_servo_signal}: the equilibrium excitation fraction changes with $\Delta$ in both the $\mathrm{HRA_p}$ \& $\mathrm{HRB_n}$ and the $\mathrm{HRA_n}$ \& $\mathrm{HRB_p}$ modified hyper-Ramsey locks, but these changes are equal and opposite. In the main part of the figure we take the difference between the equilibrium excitations fractions $1/2(P_\mathrm{HRA_n} + P_\mathrm{HRB_p}) - 1/2(P_\mathrm{HRA_p}+P_\mathrm{HRB_n})$ which can clearly be used as an error signal to steer the compensation step $\Delta_\mathrm{st}$ towards $\Delta_\mathrm{sh}$.

Further to the theoretical models, we have implemented the modified hyper-Ramsey scheme on an $^{88}$Sr lattice clock to gather data plotted in figures \ref{fig:Hyper-Ramsey_shift_curve} and \ref{fig:Excitation-based_step_servo_signal}. For this data we employ a similar cooling sequence as previously used in other $^{88}$Sr systems \cite{Akatsuka2010,Lisdat2009}. In brief, we begin by capturing atoms exiting from a permanent-magnet Zeeman slower \cite{Hill2014} into a `blue' magneto-optical trap (MOT) operating on the broad $\SI{461}{\nano\meter}$ $^1$S$_0$ $-$ $^1$P$_1$ transition. After $\SI{60}{\milli\second}$ the slowing beam is switched off and the blue MOT light is ramped down to zero intensity in $\SI{10}{\milli\second}$ in order to allow the cold atoms to move closer to the center of the $\SI{4}{\milli\tesla\per\centi\meter}$ quadrupole magnetic field. Next, the atoms are transferred into a broadband `red' MOT operating on the narrow $\SI{689}{\nano\meter}$ $^1$S$_0$ $-$ $^3$P$_1$ transition. For the red MOT the quadrupole field is quickly switched to $\SI{0.2}{\milli\tesla\per\centi\meter}$ and the light is modulated to cover a $\SI{4}{\mega\hertz}$ spectral region, ensuring resonance with all but the fastest atoms left over in the blue MOT. After $\SI{60}{\milli\second}$ in the broadband red MOT, the modulation is switched off to realize the $\sim \SI{1}{\micro\kelvin}$ temperature required to load efficiently into a one-dimensional vertically-oriented optical lattice. The lattice has a $\SI{45}{\micro\meter}$ waist and is formed by retro-reflection of up to $\SI{850}{\milli\watt}$ of light from a Ti:Sapphire laser operating near the $^{88}$Sr magic wavelength at $\SI{368554.5}{\giga\hertz}$ \cite{Akatsuka2010}. The last stage of narrow-band cooling in the red MOT lasts a total of $\SI{120}{\milli\second}$, and in the latter half of this time we ramp the $\SI{689}{\nano\meter}$ detuning by $-\SI{200}{\kilo\hertz}$ to allow the MOT to fall under gravity in a controlled manner, loading the lattice along a $\SI{300}{\micro\meter}$ streak covering around 700 lattice sites. Before interrogating the clock transition we ramp the lattice depth from $\SI{11}{\micro\kelvin}$ to around $\SI{2}{\micro\kelvin}$ in $\SI{30}{\milli\second}$, hold for $\SI{30}{\milli\second}$, and then ramp back up to $\SI{11}{\micro\kelvin}$ in another $\SI{30}{\milli\second}$. This stage ensures that the hotter atoms are discarded and gives a stable method to load at low atomic density.

We interrogate the $\SI{698}{\nano\meter}$ $^1$S$_0$ $-$ $^3$P$_0$ clock transition in $^{88}$Sr using magnetically-induced spectroscopy \cite{Taichenachev2006a}. The clock light is sourced from a slave laser diode injected by the transmission of a high-finesse vertically-oriented reference cavity \cite{Ludlow2007}. In a scheme bearing some similarity with \cite{Hagemann2013}, extra frequency stability is fed forward onto the interrogating light with $\SI{30}{\hertz}$ bandwidth using a fiber-comb-derived transfer beat \cite{Telle2001} against a more stable laser at $\SI{1064}{\nano\meter}$, enabling coherent interrogation of atomic features down to a $\SI{2}{\hertz}$ linewidth. We focus $\SI{2.7}{\milli\watt}$ of interrogating light to a waist of approximately $\SI{250}{\micro\meter}$, combined with the optical lattice light on a dichroic beamsplitter. In order to compensate for AOM chirp and fibre phase noise the clock delivery is actively phase-stabilized \cite{Ma1994}, with constant compensation of the final `switching' AOM achieved by retro-reflecting the zeroth-order beam. The magnetic field required to access the clock transition is applied using a pair of coils in near-Helmholtz configuration, with good field stability ensured by active stabilization to a high-specification current transducer. The mixing field was fixed at $\SI{2.5}{\milli\tesla}$ throughout all interrogation sequences used in this paper.

For the data in figure \ref{fig:Hyper-Ramsey_shift_curve} we operate with 4 active clock servos. Servo 1 is the reference, using a modified hyper-Ramsey $\mathrm{HRA_p}$ \& $\mathrm{HRB_n}$ sequence with a compensation step of $\Delta_\mathrm{st} = \SI{80}{\hertz}$ close to the slowly-varying probe Stark shift $79.6 < \Delta_\mathrm{sh}/2\pi < \SI{80.7}{\hertz}$. Servo 2 then uses one of the two types of hyper-Ramsey sequence investigated in the figure, applying various different compensation steps $66 < \Delta_\mathrm{st}/2\pi < \SI{94}{\hertz}$ to map out the residual Stark shift. The $y$-position of plotted points is given by the frequency offset between servos 1 and 2. Meanwhile, servos 3 and 4 utilize Rabi interrogation as a live monitor the Stark shift. They are set to run in only 4 out of 16 cycles to avoid degrading the stability of the hyper-Ramsey servos. Servo 3 uses the same intensity and magnetic field as the hyper-Ramsey servos and is therefore shifted by $\Delta_\mathrm{sh}$ from servo 1, allowing us to calculate the $x$-position of the plotted points in the figure. Servo 4 uses a lower intensity but the same field as servo 3, allowing us to extrapolate an independent (but less accurate) measure of $\Delta_\mathrm{sh}$. For the data in figure \ref{fig:Excitation-based_step_servo_signal} we use the same set of 4 servos except that the compensation step in servo 1 is no longer set $\SI{80}{\hertz}$ but is instead scanned in tandem with that of servo 2.

To conclude, the data in figure \ref{fig:Hyper-Ramsey_shift_curve} verifies our most important result: within the \num{1e-16} measurement statistics the modified hyper-Ramsey scheme proves completely immune to the \num{2e-13} probe Stark shift over a large range of compensation steps. In theory the modified hyper-Ramsey scheme should have no probe shift at all, but at much lower levels of uncertainty we anticipate two potential sources for residual lock offset going forward: (1) sampling of asymmetry in the error signal due to occasional large clock laser deviations, and (2) imperfect realization of the hyper-Ramsey pulse due to residual clock phase noise or motional heating of the atoms. In the optical lattice clock we expect these effects to be controllable well below the $10^{-18}$ level, though the much higher heating rates in typical ion traps may mean that more care is required to reach the same uncertainty in ion-based systems. In future work, we hope to apply the modified hyper-Ramsey scheme to suppress the $\SI{150}{\hertz}$ quadratic Zeeman shift as well as the probe Stark shift, eliminating a large part of the $^{88}$Sr uncertainty budget.

We thank Helen Margolis, Fred Baynes and Antoine Rolland for providing us with a transfer beat for clock laser stabilization, and Ross Williams for setting up the $\SI{1064}{\nano\meter}$ cavity. We are grateful for stimulating discussions with Peter Nisbet-Jones, Rachel Godun, Charles Baynham and Jonathan Jones. This work was funded by the UK Department for Business, Innovation and Skills, as part of the National Measurement System Electromagnetics and Time programme.

\bibliography{../library1}

\begin{thebibliography}{29}
\expandafter\ifx\csname natexlab\endcsname\relax\def\natexlab#1{#1}\fi
\expandafter\ifx\csname bibnamefont\endcsname\relax
  \def\bibnamefont#1{#1}\fi
\expandafter\ifx\csname bibfnamefont\endcsname\relax
  \def\bibfnamefont#1{#1}\fi
\expandafter\ifx\csname citenamefont\endcsname\relax
  \def\citenamefont#1{#1}\fi
\expandafter\ifx\csname url\endcsname\relax
  \def\url#1{\texttt{#1}}\fi
\expandafter\ifx\csname urlprefix\endcsname\relax\def\urlprefix{URL }\fi
\providecommand{\bibinfo}[2]{#2}
\providecommand{\eprint}[2][]{\url{#2}}

\bibitem[{\citenamefont{Hinkley et~al.}(2013)\citenamefont{Hinkley, Sherman,
  Phillips, Schioppo, Lemke, Beloy, Pizzocaro, Oates, and
  Ludlow}}]{Hinkley2013}
\bibinfo{author}{\bibfnamefont{N.}~\bibnamefont{Hinkley}},
  \bibinfo{author}{\bibfnamefont{J.~A.} \bibnamefont{Sherman}},
  \bibinfo{author}{\bibfnamefont{N.~B.} \bibnamefont{Phillips}},
  \bibinfo{author}{\bibfnamefont{M.}~\bibnamefont{Schioppo}},
  \bibinfo{author}{\bibfnamefont{N.~D.} \bibnamefont{Lemke}},
  \bibinfo{author}{\bibfnamefont{K.}~\bibnamefont{Beloy}},
  \bibinfo{author}{\bibfnamefont{M.}~\bibnamefont{Pizzocaro}},
  \bibinfo{author}{\bibfnamefont{C.~W.} \bibnamefont{Oates}}, \bibnamefont{and}
  \bibinfo{author}{\bibfnamefont{A.~D.} \bibnamefont{Ludlow}},
  \bibinfo{journal}{Science}  (\bibinfo{year}{2013}),
  \urlprefix\url{http://www.sciencemag.org/content/early/2013/08/21/science.1240420.abstract}.

\bibitem[{\citenamefont{Nicholson et~al.}(2015)\citenamefont{Nicholson,
  Campbell, Hutson, Marti, Bloom, McNally, Zhang, Barrett, Safronova, Strouse
  et~al.}}]{Nicholson2015}
\bibinfo{author}{\bibfnamefont{T.~L.} \bibnamefont{Nicholson}},
  \bibinfo{author}{\bibfnamefont{S.~L.} \bibnamefont{Campbell}},
  \bibinfo{author}{\bibfnamefont{R.~B.} \bibnamefont{Hutson}},
  \bibinfo{author}{\bibfnamefont{G.~E.} \bibnamefont{Marti}},
  \bibinfo{author}{\bibfnamefont{B.~J.} \bibnamefont{Bloom}},
  \bibinfo{author}{\bibfnamefont{R.~L.} \bibnamefont{McNally}},
  \bibinfo{author}{\bibfnamefont{W.}~\bibnamefont{Zhang}},
  \bibinfo{author}{\bibfnamefont{M.~D.} \bibnamefont{Barrett}},
  \bibinfo{author}{\bibfnamefont{M.~S.} \bibnamefont{Safronova}},
  \bibinfo{author}{\bibfnamefont{G.~F.} \bibnamefont{Strouse}},
  \bibnamefont{et~al.}, \bibinfo{journal}{Nat Commun}
  \textbf{\bibinfo{volume}{6}} (\bibinfo{year}{2015}),
  \urlprefix\url{http://dx.doi.org/10.1038/ncomms7896}.

\bibitem[{\citenamefont{Gill}(2011)}]{Gill2011}
\bibinfo{author}{\bibfnamefont{P.}~\bibnamefont{Gill}},
  \bibinfo{journal}{Philosophical Transactions of the Royal Society of London
  A: Mathematical, Physical and Engineering Sciences}
  \textbf{\bibinfo{volume}{369}}, \bibinfo{pages}{4109} (\bibinfo{year}{2011}),
  ISSN \bibinfo{issn}{1364-503X}.

\bibitem[{\citenamefont{Chou et~al.}(2010)\citenamefont{Chou, Hume, Koelemeij,
  Wineland, and Rosenband}}]{Chou2010}
\bibinfo{author}{\bibfnamefont{C.~W.} \bibnamefont{Chou}},
  \bibinfo{author}{\bibfnamefont{D.~B.} \bibnamefont{Hume}},
  \bibinfo{author}{\bibfnamefont{J.~C.~J.} \bibnamefont{Koelemeij}},
  \bibinfo{author}{\bibfnamefont{D.~J.} \bibnamefont{Wineland}},
  \bibnamefont{and}
  \bibinfo{author}{\bibfnamefont{T.}~\bibnamefont{Rosenband}},
  \bibinfo{journal}{Phys. Rev. Lett.} \textbf{\bibinfo{volume}{104}},
  \bibinfo{pages}{070802} (\bibinfo{year}{2010}),
  \urlprefix\url{http://link.aps.org/doi/10.1103/PhysRevLett.104.070802}.

\bibitem[{\citenamefont{Derevianko and Pospelov}(2014)}]{Derevianko2014}
\bibinfo{author}{\bibfnamefont{A.}~\bibnamefont{Derevianko}} \bibnamefont{and}
  \bibinfo{author}{\bibfnamefont{M.}~\bibnamefont{Pospelov}},
  \bibinfo{journal}{Nat Phys} \textbf{\bibinfo{volume}{10}},
  \bibinfo{pages}{933} (\bibinfo{year}{2014}), ISSN \bibinfo{issn}{1745-2473},
  \bibinfo{note}{letter}, \urlprefix\url{http://dx.doi.org/10.1038/nphys3137}.

\bibitem[{\citenamefont{Van~Tilburg et~al.}(2015)\citenamefont{Van~Tilburg,
  Leefer, Bougas, and Budker}}]{VanTilburg2015}
\bibinfo{author}{\bibfnamefont{K.}~\bibnamefont{Van~Tilburg}},
  \bibinfo{author}{\bibfnamefont{N.}~\bibnamefont{Leefer}},
  \bibinfo{author}{\bibfnamefont{L.}~\bibnamefont{Bougas}}, \bibnamefont{and}
  \bibinfo{author}{\bibfnamefont{D.}~\bibnamefont{Budker}},
  \bibinfo{journal}{Phys. Rev. Lett.} \textbf{\bibinfo{volume}{115}},
  \bibinfo{pages}{011802} (\bibinfo{year}{2015}),
  \urlprefix\url{http://link.aps.org/doi/10.1103/PhysRevLett.115.011802}.

\bibitem[{\citenamefont{Godun et~al.}(2014)\citenamefont{Godun, Nisbet-Jones,
  Jones, King, Johnson, Margolis, Szymaniec, Lea, Bongs, and Gill}}]{Godun2014}
\bibinfo{author}{\bibfnamefont{R.~M.} \bibnamefont{Godun}},
  \bibinfo{author}{\bibfnamefont{P.~B.~R.} \bibnamefont{Nisbet-Jones}},
  \bibinfo{author}{\bibfnamefont{J.~M.} \bibnamefont{Jones}},
  \bibinfo{author}{\bibfnamefont{S.~A.} \bibnamefont{King}},
  \bibinfo{author}{\bibfnamefont{L.~A.~M.} \bibnamefont{Johnson}},
  \bibinfo{author}{\bibfnamefont{H.~S.} \bibnamefont{Margolis}},
  \bibinfo{author}{\bibfnamefont{K.}~\bibnamefont{Szymaniec}},
  \bibinfo{author}{\bibfnamefont{S.~N.} \bibnamefont{Lea}},
  \bibinfo{author}{\bibfnamefont{K.}~\bibnamefont{Bongs}}, \bibnamefont{and}
  \bibinfo{author}{\bibfnamefont{P.}~\bibnamefont{Gill}},
  \bibinfo{journal}{Phys. Rev. Lett.} \textbf{\bibinfo{volume}{113}},
  \bibinfo{pages}{210801} (\bibinfo{year}{2014}),
  \urlprefix\url{http://link.aps.org/doi/10.1103/PhysRevLett.113.210801}.

\bibitem[{\citenamefont{Huntemann et~al.}(2014)\citenamefont{Huntemann,
  Lipphardt, Tamm, Gerginov, Weyers, and Peik}}]{Huntemann2014}
\bibinfo{author}{\bibfnamefont{N.}~\bibnamefont{Huntemann}},
  \bibinfo{author}{\bibfnamefont{B.}~\bibnamefont{Lipphardt}},
  \bibinfo{author}{\bibfnamefont{C.}~\bibnamefont{Tamm}},
  \bibinfo{author}{\bibfnamefont{V.}~\bibnamefont{Gerginov}},
  \bibinfo{author}{\bibfnamefont{S.}~\bibnamefont{Weyers}}, \bibnamefont{and}
  \bibinfo{author}{\bibfnamefont{E.}~\bibnamefont{Peik}},
  \bibinfo{journal}{Phys. Rev. Lett.} \textbf{\bibinfo{volume}{113}},
  \bibinfo{pages}{210802} (\bibinfo{year}{2014}),
  \urlprefix\url{http://link.aps.org/doi/10.1103/PhysRevLett.113.210802}.

\bibitem[{\citenamefont{Huntemann
  et~al.}(2012{\natexlab{a}})\citenamefont{Huntemann, Okhapkin, Lipphardt,
  Weyers, Tamm, and Peik}}]{Huntemann2012}
\bibinfo{author}{\bibfnamefont{N.}~\bibnamefont{Huntemann}},
  \bibinfo{author}{\bibfnamefont{M.}~\bibnamefont{Okhapkin}},
  \bibinfo{author}{\bibfnamefont{B.}~\bibnamefont{Lipphardt}},
  \bibinfo{author}{\bibfnamefont{S.}~\bibnamefont{Weyers}},
  \bibinfo{author}{\bibfnamefont{C.}~\bibnamefont{Tamm}}, \bibnamefont{and}
  \bibinfo{author}{\bibfnamefont{E.}~\bibnamefont{Peik}},
  \bibinfo{journal}{Phys. Rev. Lett.} \textbf{\bibinfo{volume}{108}},
  \bibinfo{pages}{090801} (\bibinfo{year}{2012}{\natexlab{a}}),
  \urlprefix\url{http://link.aps.org/doi/10.1103/PhysRevLett.108.090801}.

\bibitem[{\citenamefont{King et~al.}(2012)\citenamefont{King, Godun, Webster,
  Margolis, Johnson, Szymaniec, Baird, and Gill}}]{King2012}
\bibinfo{author}{\bibfnamefont{S.}~\bibnamefont{King}},
  \bibinfo{author}{\bibfnamefont{R.}~\bibnamefont{Godun}},
  \bibinfo{author}{\bibfnamefont{S.}~\bibnamefont{Webster}},
  \bibinfo{author}{\bibfnamefont{H.}~\bibnamefont{Margolis}},
  \bibinfo{author}{\bibfnamefont{L.}~\bibnamefont{Johnson}},
  \bibinfo{author}{\bibfnamefont{K.}~\bibnamefont{Szymaniec}},
  \bibinfo{author}{\bibfnamefont{P.}~\bibnamefont{Baird}}, \bibnamefont{and}
  \bibinfo{author}{\bibfnamefont{P.}~\bibnamefont{Gill}}, \bibinfo{journal}{New
  Journal of Physics} \textbf{\bibinfo{volume}{14}}, \bibinfo{pages}{013045}
  (\bibinfo{year}{2012}),
  \urlprefix\url{http://stacks.iop.org/1367-2630/14/i=1/a=013045}.

\bibitem[{\citenamefont{Zanon-Willette
  et~al.}(2006)\citenamefont{Zanon-Willette, Ludlow, Blatt, Boyd, Arimondo, and
  Ye}}]{Zanon-Willette2006}
\bibinfo{author}{\bibfnamefont{T.}~\bibnamefont{Zanon-Willette}},
  \bibinfo{author}{\bibfnamefont{A.~D.} \bibnamefont{Ludlow}},
  \bibinfo{author}{\bibfnamefont{S.}~\bibnamefont{Blatt}},
  \bibinfo{author}{\bibfnamefont{M.~M.} \bibnamefont{Boyd}},
  \bibinfo{author}{\bibfnamefont{E.}~\bibnamefont{Arimondo}}, \bibnamefont{and}
  \bibinfo{author}{\bibfnamefont{J.}~\bibnamefont{Ye}}, \bibinfo{journal}{Phys.
  Rev. Lett.} \textbf{\bibinfo{volume}{97}}, \bibinfo{pages}{233001}
  (\bibinfo{year}{2006}),
  \urlprefix\url{http://link.aps.org/doi/10.1103/PhysRevLett.97.233001}.

\bibitem[{\citenamefont{Parthey et~al.}(2011)\citenamefont{Parthey, Matveev,
  Alnis, Bernhardt, Beyer, Holzwarth, Maistrou, Pohl, Predehl, Udem
  et~al.}}]{Parthey2011}
\bibinfo{author}{\bibfnamefont{C.~G.} \bibnamefont{Parthey}},
  \bibinfo{author}{\bibfnamefont{A.}~\bibnamefont{Matveev}},
  \bibinfo{author}{\bibfnamefont{J.}~\bibnamefont{Alnis}},
  \bibinfo{author}{\bibfnamefont{B.}~\bibnamefont{Bernhardt}},
  \bibinfo{author}{\bibfnamefont{A.}~\bibnamefont{Beyer}},
  \bibinfo{author}{\bibfnamefont{R.}~\bibnamefont{Holzwarth}},
  \bibinfo{author}{\bibfnamefont{A.}~\bibnamefont{Maistrou}},
  \bibinfo{author}{\bibfnamefont{R.}~\bibnamefont{Pohl}},
  \bibinfo{author}{\bibfnamefont{K.}~\bibnamefont{Predehl}},
  \bibinfo{author}{\bibfnamefont{T.}~\bibnamefont{Udem}}, \bibnamefont{et~al.},
  \bibinfo{journal}{Phys. Rev. Lett.} \textbf{\bibinfo{volume}{107}},
  \bibinfo{pages}{203001} (\bibinfo{year}{2011}),
  \urlprefix\url{http://link.aps.org/doi/10.1103/PhysRevLett.107.203001}.

\bibitem[{\citenamefont{Taichenachev et~al.}(2006)\citenamefont{Taichenachev,
  Yudin, Oates, Hoyt, Barber, and Hollberg}}]{Taichenachev2006a}
\bibinfo{author}{\bibfnamefont{A.~V.} \bibnamefont{Taichenachev}},
  \bibinfo{author}{\bibfnamefont{V.~I.} \bibnamefont{Yudin}},
  \bibinfo{author}{\bibfnamefont{C.~W.} \bibnamefont{Oates}},
  \bibinfo{author}{\bibfnamefont{C.~W.} \bibnamefont{Hoyt}},
  \bibinfo{author}{\bibfnamefont{Z.~W.} \bibnamefont{Barber}},
  \bibnamefont{and} \bibinfo{author}{\bibfnamefont{L.}~\bibnamefont{Hollberg}},
  \bibinfo{journal}{Phys. Rev. Lett.} \textbf{\bibinfo{volume}{96}},
  \bibinfo{pages}{083001} (\bibinfo{year}{2006}),
  \urlprefix\url{http://link.aps.org/doi/10.1103/PhysRevLett.96.083001}.

\bibitem[{\citenamefont{Barber et~al.}(2006)\citenamefont{Barber, Hoyt, Oates,
  Hollberg, Taichenachev, and Yudin}}]{Barber2006}
\bibinfo{author}{\bibfnamefont{Z.~W.} \bibnamefont{Barber}},
  \bibinfo{author}{\bibfnamefont{C.~W.} \bibnamefont{Hoyt}},
  \bibinfo{author}{\bibfnamefont{C.~W.} \bibnamefont{Oates}},
  \bibinfo{author}{\bibfnamefont{L.}~\bibnamefont{Hollberg}},
  \bibinfo{author}{\bibfnamefont{A.~V.} \bibnamefont{Taichenachev}},
  \bibnamefont{and} \bibinfo{author}{\bibfnamefont{V.~I.} \bibnamefont{Yudin}},
  \bibinfo{journal}{Phys. Rev. Lett.} \textbf{\bibinfo{volume}{96}},
  \bibinfo{pages}{083002} (\bibinfo{year}{2006}),
  \urlprefix\url{http://link.aps.org/doi/10.1103/PhysRevLett.96.083002}.

\bibitem[{\citenamefont{Akatsuka et~al.}(2010)\citenamefont{Akatsuka, Takamoto,
  and Katori}}]{Akatsuka2010}
\bibinfo{author}{\bibfnamefont{T.}~\bibnamefont{Akatsuka}},
  \bibinfo{author}{\bibfnamefont{M.}~\bibnamefont{Takamoto}}, \bibnamefont{and}
  \bibinfo{author}{\bibfnamefont{H.}~\bibnamefont{Katori}},
  \bibinfo{journal}{Phys. Rev. A} \textbf{\bibinfo{volume}{81}},
  \bibinfo{pages}{023402} (\bibinfo{year}{2010}),
  \urlprefix\url{http://link.aps.org/doi/10.1103/PhysRevA.81.023402}.

\bibitem[{\citenamefont{Taichenachev et~al.}(2010)\citenamefont{Taichenachev,
  Yudin, Oates, Barber, Lemke, Ludlow, Sterr, Lisdat, and
  Riehle}}]{Taichenachev2010}
\bibinfo{author}{\bibfnamefont{A.}~\bibnamefont{Taichenachev}},
  \bibinfo{author}{\bibfnamefont{V.}~\bibnamefont{Yudin}},
  \bibinfo{author}{\bibfnamefont{C.}~\bibnamefont{Oates}},
  \bibinfo{author}{\bibfnamefont{Z.}~\bibnamefont{Barber}},
  \bibinfo{author}{\bibfnamefont{N.}~\bibnamefont{Lemke}},
  \bibinfo{author}{\bibfnamefont{A.}~\bibnamefont{Ludlow}},
  \bibinfo{author}{\bibfnamefont{U.}~\bibnamefont{Sterr}},
  \bibinfo{author}{\bibfnamefont{C.}~\bibnamefont{Lisdat}}, \bibnamefont{and}
  \bibinfo{author}{\bibfnamefont{F.}~\bibnamefont{Riehle}},
  \bibinfo{journal}{JETP Letters} \textbf{\bibinfo{volume}{90}},
  \bibinfo{pages}{713} (\bibinfo{year}{2010}), ISSN \bibinfo{issn}{0021-3640},
  \urlprefix\url{http://dx.doi.org/10.1134/S0021364009230052}.

\bibitem[{\citenamefont{Yudin et~al.}(2010)\citenamefont{Yudin, Taichenachev,
  Oates, Barber, Lemke, Ludlow, Sterr, Lisdat, and Riehle}}]{Yudin2010}
\bibinfo{author}{\bibfnamefont{V.~I.} \bibnamefont{Yudin}},
  \bibinfo{author}{\bibfnamefont{A.~V.} \bibnamefont{Taichenachev}},
  \bibinfo{author}{\bibfnamefont{C.~W.} \bibnamefont{Oates}},
  \bibinfo{author}{\bibfnamefont{Z.~W.} \bibnamefont{Barber}},
  \bibinfo{author}{\bibfnamefont{N.~D.} \bibnamefont{Lemke}},
  \bibinfo{author}{\bibfnamefont{A.~D.} \bibnamefont{Ludlow}},
  \bibinfo{author}{\bibfnamefont{U.}~\bibnamefont{Sterr}},
  \bibinfo{author}{\bibfnamefont{C.}~\bibnamefont{Lisdat}}, \bibnamefont{and}
  \bibinfo{author}{\bibfnamefont{F.}~\bibnamefont{Riehle}},
  \bibinfo{journal}{Phys. Rev. A} \textbf{\bibinfo{volume}{82}},
  \bibinfo{pages}{011804} (\bibinfo{year}{2010}),
  \urlprefix\url{http://link.aps.org/doi/10.1103/PhysRevA.82.011804}.

\bibitem[{\citenamefont{Zanon-Willette
  et~al.}(2014)\citenamefont{Zanon-Willette, Almonacil, de~Clercq, Ludlow, and
  Arimondo}}]{Zanon-Willette2014}
\bibinfo{author}{\bibfnamefont{T.}~\bibnamefont{Zanon-Willette}},
  \bibinfo{author}{\bibfnamefont{S.}~\bibnamefont{Almonacil}},
  \bibinfo{author}{\bibfnamefont{E.}~\bibnamefont{de~Clercq}},
  \bibinfo{author}{\bibfnamefont{A.~D.} \bibnamefont{Ludlow}},
  \bibnamefont{and} \bibinfo{author}{\bibfnamefont{E.}~\bibnamefont{Arimondo}},
  \bibinfo{journal}{Phys. Rev. A} \textbf{\bibinfo{volume}{90}},
  \bibinfo{pages}{053427} (\bibinfo{year}{2014}),
  \urlprefix\url{http://link.aps.org/doi/10.1103/PhysRevA.90.053427}.

\bibitem[{\citenamefont{Zanon-Willette
  et~al.}(2015)\citenamefont{Zanon-Willette, Yudin, and
  Taichenachev}}]{Zanon-Willette2015}
\bibinfo{author}{\bibfnamefont{T.}~\bibnamefont{Zanon-Willette}},
  \bibinfo{author}{\bibfnamefont{V.~I.} \bibnamefont{Yudin}}, \bibnamefont{and}
  \bibinfo{author}{\bibfnamefont{A.~V.} \bibnamefont{Taichenachev}},
  \bibinfo{journal}{Phys. Rev. A} \textbf{\bibinfo{volume}{92}},
  \bibinfo{pages}{023416} (\bibinfo{year}{2015}),
  \urlprefix\url{http://link.aps.org/doi/10.1103/PhysRevA.92.023416}.

\bibitem[{\citenamefont{Huntemann
  et~al.}(2012{\natexlab{b}})\citenamefont{Huntemann, Lipphardt, Okhapkin,
  Tamm, Peik, Taichenachev, and Yudin}}]{Huntemann2012a}
\bibinfo{author}{\bibfnamefont{N.}~\bibnamefont{Huntemann}},
  \bibinfo{author}{\bibfnamefont{B.}~\bibnamefont{Lipphardt}},
  \bibinfo{author}{\bibfnamefont{M.}~\bibnamefont{Okhapkin}},
  \bibinfo{author}{\bibfnamefont{C.}~\bibnamefont{Tamm}},
  \bibinfo{author}{\bibfnamefont{E.}~\bibnamefont{Peik}},
  \bibinfo{author}{\bibfnamefont{A.~V.} \bibnamefont{Taichenachev}},
  \bibnamefont{and} \bibinfo{author}{\bibfnamefont{V.~I.} \bibnamefont{Yudin}},
  \bibinfo{journal}{Phys. Rev. Lett.} \textbf{\bibinfo{volume}{109}},
  \bibinfo{pages}{213002} (\bibinfo{year}{2012}{\natexlab{b}}),
  \urlprefix\url{http://link.aps.org/doi/10.1103/PhysRevLett.109.213002}.

\bibitem[{\citenamefont{Fortier et~al.}(2006)\citenamefont{Fortier, Le~Coq,
  Stalnaker, Ortega, Diddams, Oates, and Hollberg}}]{Fortier2006}
\bibinfo{author}{\bibfnamefont{T.~M.} \bibnamefont{Fortier}},
  \bibinfo{author}{\bibfnamefont{Y.}~\bibnamefont{Le~Coq}},
  \bibinfo{author}{\bibfnamefont{J.~E.} \bibnamefont{Stalnaker}},
  \bibinfo{author}{\bibfnamefont{D.}~\bibnamefont{Ortega}},
  \bibinfo{author}{\bibfnamefont{S.~A.} \bibnamefont{Diddams}},
  \bibinfo{author}{\bibfnamefont{C.~W.} \bibnamefont{Oates}}, \bibnamefont{and}
  \bibinfo{author}{\bibfnamefont{L.}~\bibnamefont{Hollberg}},
  \bibinfo{journal}{Phys. Rev. Lett.} \textbf{\bibinfo{volume}{97}},
  \bibinfo{pages}{163905} (\bibinfo{year}{2006}),
  \urlprefix\url{http://link.aps.org/doi/10.1103/PhysRevLett.97.163905}.

\bibitem[{\citenamefont{Safronova et~al.}(2014)\citenamefont{Safronova, Dzuba,
  Flambaum, Safronova, Porsev, and Kozlov}}]{Safronova2014}
\bibinfo{author}{\bibfnamefont{M.~S.} \bibnamefont{Safronova}},
  \bibinfo{author}{\bibfnamefont{V.~A.} \bibnamefont{Dzuba}},
  \bibinfo{author}{\bibfnamefont{V.~V.} \bibnamefont{Flambaum}},
  \bibinfo{author}{\bibfnamefont{U.~I.} \bibnamefont{Safronova}},
  \bibinfo{author}{\bibfnamefont{S.~G.} \bibnamefont{Porsev}},
  \bibnamefont{and} \bibinfo{author}{\bibfnamefont{M.~G.}
  \bibnamefont{Kozlov}}, \bibinfo{journal}{Phys. Rev. Lett.}
  \textbf{\bibinfo{volume}{113}}, \bibinfo{pages}{030801}
  (\bibinfo{year}{2014}),
  \urlprefix\url{http://link.aps.org/doi/10.1103/PhysRevLett.113.030801}.

\bibitem[{\citenamefont{Ramsey}(1950)}]{Ramsey1950}
\bibinfo{author}{\bibfnamefont{N.}~\bibnamefont{Ramsey}},
  \bibinfo{journal}{Physical Review} \textbf{\bibinfo{volume}{78}},
  \bibinfo{pages}{695} (\bibinfo{year}{1950}), ISSN \bibinfo{issn}{0031-899X},
  \urlprefix\url{http://link.aps.org/doi/10.1103/PhysRev.78.695}.

\bibitem[{\citenamefont{Lisdat et~al.}(2009)\citenamefont{Lisdat, Winfred,
  Middelmann, Riehle, and Sterr}}]{Lisdat2009}
\bibinfo{author}{\bibfnamefont{C.}~\bibnamefont{Lisdat}},
  \bibinfo{author}{\bibfnamefont{J.~S. R.~V.} \bibnamefont{Winfred}},
  \bibinfo{author}{\bibfnamefont{T.}~\bibnamefont{Middelmann}},
  \bibinfo{author}{\bibfnamefont{F.}~\bibnamefont{Riehle}}, \bibnamefont{and}
  \bibinfo{author}{\bibfnamefont{U.}~\bibnamefont{Sterr}},
  \bibinfo{journal}{Phys. Rev. Lett.} \textbf{\bibinfo{volume}{103}},
  \bibinfo{pages}{090801} (\bibinfo{year}{2009}),
  \urlprefix\url{http://link.aps.org/doi/10.1103/PhysRevLett.103.090801}.

\bibitem[{\citenamefont{Hill et~al.}(2014)\citenamefont{Hill, Ovchinnikov,
  Bridge, Curtis, and Gill}}]{Hill2014}
\bibinfo{author}{\bibfnamefont{I.~R.} \bibnamefont{Hill}},
  \bibinfo{author}{\bibfnamefont{Y.~B.} \bibnamefont{Ovchinnikov}},
  \bibinfo{author}{\bibfnamefont{E.~M.} \bibnamefont{Bridge}},
  \bibinfo{author}{\bibfnamefont{E.~A.} \bibnamefont{Curtis}},
  \bibnamefont{and} \bibinfo{author}{\bibfnamefont{P.}~\bibnamefont{Gill}},
  \bibinfo{journal}{Journal of Physics B: Atomic, Molecular and Optical
  Physics} \textbf{\bibinfo{volume}{47}}, \bibinfo{pages}{075006}
  (\bibinfo{year}{2014}),
  \urlprefix\url{http://stacks.iop.org/0953-4075/47/i=7/a=075006}.

\bibitem[{\citenamefont{Ludlow et~al.}(2007)\citenamefont{Ludlow, Huang,
  Notcutt, Zanon-Willette, Foreman, Boyd, Blatt, and Ye}}]{Ludlow2007}
\bibinfo{author}{\bibfnamefont{A.}~\bibnamefont{Ludlow}},
  \bibinfo{author}{\bibfnamefont{X.}~\bibnamefont{Huang}},
  \bibinfo{author}{\bibfnamefont{M.}~\bibnamefont{Notcutt}},
  \bibinfo{author}{\bibfnamefont{T.}~\bibnamefont{Zanon-Willette}},
  \bibinfo{author}{\bibfnamefont{S.}~\bibnamefont{Foreman}},
  \bibinfo{author}{\bibfnamefont{M.}~\bibnamefont{Boyd}},
  \bibinfo{author}{\bibfnamefont{S.}~\bibnamefont{Blatt}}, \bibnamefont{and}
  \bibinfo{author}{\bibfnamefont{J.}~\bibnamefont{Ye}},
  \bibinfo{journal}{Optics Letters} \textbf{\bibinfo{volume}{32}},
  \bibinfo{pages}{641} (\bibinfo{year}{2007}), ISSN \bibinfo{issn}{0146-9592},
  \urlprefix\url{http://www.ncbi.nlm.nih.gov/pubmed/17308587}.

\bibitem[{\citenamefont{Hagemann et~al.}(2013)\citenamefont{Hagemann, Grebing,
  Kessler, Falke, Lemke, Lisdat, Schnatz, Riehle, and Sterr}}]{Hagemann2013}
\bibinfo{author}{\bibfnamefont{C.}~\bibnamefont{Hagemann}},
  \bibinfo{author}{\bibfnamefont{C.}~\bibnamefont{Grebing}},
  \bibinfo{author}{\bibfnamefont{T.}~\bibnamefont{Kessler}},
  \bibinfo{author}{\bibfnamefont{S.}~\bibnamefont{Falke}},
  \bibinfo{author}{\bibfnamefont{N.}~\bibnamefont{Lemke}},
  \bibinfo{author}{\bibfnamefont{C.}~\bibnamefont{Lisdat}},
  \bibinfo{author}{\bibfnamefont{H.}~\bibnamefont{Schnatz}},
  \bibinfo{author}{\bibfnamefont{F.}~\bibnamefont{Riehle}}, \bibnamefont{and}
  \bibinfo{author}{\bibfnamefont{U.}~\bibnamefont{Sterr}},
  \bibinfo{journal}{Instrumentation and Measurement, IEEE Transactions on}
  \textbf{\bibinfo{volume}{62}}, \bibinfo{pages}{1556} (\bibinfo{year}{2013}),
  ISSN \bibinfo{issn}{0018-9456}.

\bibitem[{\citenamefont{Telle et~al.}(2002)\citenamefont{Telle, Lipphardt, and
  Stenger}}]{Telle2001}
\bibinfo{author}{\bibfnamefont{H.}~\bibnamefont{Telle}},
  \bibinfo{author}{\bibfnamefont{B.}~\bibnamefont{Lipphardt}},
  \bibnamefont{and} \bibinfo{author}{\bibfnamefont{J.}~\bibnamefont{Stenger}},
  \bibinfo{journal}{Applied Physics B} \textbf{\bibinfo{volume}{74}},
  \bibinfo{pages}{1} (\bibinfo{year}{2002}), ISSN \bibinfo{issn}{0946-2171},
  \urlprefix\url{http://dx.doi.org/10.1007/s003400100735}.

\bibitem[{\citenamefont{Ma et~al.}(1994)\citenamefont{Ma, Jungner, Ye, and
  Hall}}]{Ma1994}
\bibinfo{author}{\bibfnamefont{L.-S.} \bibnamefont{Ma}},
  \bibinfo{author}{\bibfnamefont{P.}~\bibnamefont{Jungner}},
  \bibinfo{author}{\bibfnamefont{J.}~\bibnamefont{Ye}}, \bibnamefont{and}
  \bibinfo{author}{\bibfnamefont{J.}~\bibnamefont{Hall}},
  \bibinfo{journal}{Optics Letters} \textbf{\bibinfo{volume}{19}},
  \bibinfo{pages}{1777} (\bibinfo{year}{1994}), ISSN \bibinfo{issn}{0146-9592},
  \urlprefix\url{http://www.ncbi.nlm.nih.gov/pubmed/19855652}.

\end{thebibliography}

\end{document}